\documentstyle[12pt,aps]{revtex}
\renewcommand{\baselinestretch}{1.25}
\newcommand{\be}{\begin{equation}}
\newcommand{\ee}{\end{equation}}
\newcommand{\bee}{\begin{eqnarray}}
\newcommand{\eee}{\end{eqnarray}}

\newcommand{\dj}{d \raisebox{1.4mm}{\hspace*{-2.75mm}-}
                                \hspace*{-1.5mm}}

\begin{document}

\begin{titlepage}
\begin{center}
      {\Large \bf $\pi N \rightarrow \eta N$ and $\eta  N  \rightarrow
      \eta  N$  partial  wave  T-matrices  in a coupled, three channel
      model }
\vspace*{10mm}\\
Mijo Batini\'{c}, Ivo \v{S}laus, Alfred \v{S}varc   \\
{\em Ru\dj er Bo\v{s}kovi\'{c} Institute,  Zagreb,
 \vspace*{1.cm} Croatia } \\ and \\
 B.M.K. Nefkens \\ {\em University of California Los Angeles \\
     USA}
\end{center}
      The  $\pi  N  \rightarrow  \eta N$ and $\eta N \rightarrow \eta N$
      partial wave T-matrices for the eight lowest  partial  waves  have
      been  obtained  in  a  three  coupled channel model with unitarity
      manifestly imposed.  The two physical channels  are  $\pi  N$  and
      $\eta  N$, and the third channel, $\pi \pi N$ is an effective, but
      unphysical  two  body  channel  which  represents  all   remaining
      processes.  The $\pi N$ elastic phase shifts and the weighted data
      base of the $\pi N \rightarrow  \eta  N$  total  and  differential
      cross  sections are chosen as the input for the fitting procedure.
      A model containing a single resonance in each of the three partial
      waves  that  dominates  the $\eta$ production at lower energies is
      compared with previous analyses, based on similar assumptions.   A
      multiresonance   coupled   channel   model   is  introduced  which
      significantly improves the agreement with  all  input  data.   Our
      results  are  compared with a complementary multiresonance coupled
      channel analysis that is constrained with  elastic  and  continuum
      production   channels.   The  inclusion  of  the  fourth  P$_{11}$
      resonance in the 1440 -  2200  MeV  region  further  improves  the
      agreement between the analysis and the data.
\end{titlepage}
\renewcommand{\baselinestretch}{1.8}
\newpage
\setcounter{page}{1}
\section{Introduction}
     The cross section for $\eta$-meson production by pions
\be
   \pi^-   p  \rightarrow \eta  n
\ee
      has  a  large  peak  at  an  energy close to the $\eta$ production
      threshold, see Figs.  1a and 1b.  The maximum value, $\sigma_{tot}
      (\pi^-   p   \rightarrow   \eta n)$   =   2.8   mb,  is  6  \%  of
      $\sigma_{tot}(\pi^- p \rightarrow  {\rm  all}  )$  \cite{Pdg  92}.
      This  large  peak  is  usually  associated with the S$_{11}$(1535)
      isobar known from $\pi N$ elastic scattering and  photoproduction.
      Recently,  H\"{o}hler  \cite{Hoh  93}  has  drawn  attention  to a
      peculiar feature of this  S$_{11}$  resonance,  namely,  that  the
      speed plot analysis of S-wave $\pi N$ scattering has a sharp spike
      at the opening of the $\eta$ channel  and  no  clear,  independent
      indication  of  a  resonance in the energy region around 1535 MeV.
      This suggests a strong interplay between the cusp associated  with
      the  opening  of  the  $\eta$  channel  and  the excitation of the
      S$_{11}$ resonance.

      Some time ago Bhalerao and Liu \cite{Bha  85}  analyzed  the  then
      available  $\pi  N  \rightarrow \eta N$ data in a coupled-channel,
      single resonance separable interaction model in which the reaction
      proceeds  via  the  formation  of $\Delta$ and $N^*$ isobars - and
      concluded that the S-wave  $\eta  N$  interaction  is  attractive.
      They  found  the value of the S-wave scattering length of $a_{\eta
      N} = (0.27 + i\,0.22)$ fm.  Arima  et  al  \cite{Ari  93}  obtained
      $a_{\eta  N}  =  (0.98  +  i\,0.37)$ fm, while Wilkin, based on an
      S-wave threshold enhancement  calculation,  quotes  the  value  of
      $a_{\eta  N}  = (0.55 \pm 0.20 + i\,0.30)$ fm \cite{Wil 93}.  The
      large spread in these  values  of  the  fundamental  $a_{\eta  N}$
      parameter  illustrates  the  need  for better understanding of the
      $\eta N$ system at low energies.  The first step to do this is  to
      obtain a reliable set of $\pi N \rightarrow \eta N$ T-matrices.

      The  indication  of strong and attractive $\eta N$ interaction has
      led to a speculation about the existence of a new type of  nuclear
      matter,  namely,  quasi-bound  $\eta$-mesic  nuclei \cite{Liu 86}.
      The properties of this new matter are determined by the  $\eta  N$
      interaction at low energies.

      Good  data  on  $\eta$  production  in  $\pi^- p$   interaction is
      missing.  The dominant contribution to the surprisingly big $\eta$
      production   channel   is  coming  from  the  S$_{11}$(1535)
      resonance,   the   contribution   of   the   P$_{11}$(1440)    and
      D$_{13}$(1520)   resonances   is  important,  but  not  completely
      clarified.  The role of other resonances, even  in  these  partial
      waves,  is  not  at  all discussed because of the single resonance
      character of the model.  Recently, accurate $\eta$ photoproduction
      data  have  been  obtained  by  TAPS  at  MAMI \cite{Kru 93} up to
      $E_{\gamma}$ = 790 MeV.  These data  indicate  that  the  D$_{13}$
      resonance contribution is small.

      The  objective  of  this  paper  is  to  furnish  a  set of $\pi N
      \rightarrow \eta N$ partial wave  T  matrices  that  describe  the
      available  data  in a straightforward way.  It is essential to get
      reliable information on $\pi N \rightarrow  \eta  N$  on  shell  T
      matrices  to be able to calculate the higher order processes where
      the $\eta$ production  vertex  is  a  part  of  the  higher  order
      diagram,  for  instance in  $p\,p \rightarrow p\,p\,\eta$. Several
      publications, dealing with higher  order  processes,  have  either
      used  the  dominant  S$_{11}$ partial wave \cite{Vet 91,Lag 91,Car
      93} or have included other partial waves \cite{Bha 85,Chi 91}  for
      obtaining  the  elementary  $\pi N \rightarrow \eta N$ amplitudes.
      It is clear from Fig.  1a and 1b, where a comparison of  different
      model  predictions  with experimental data is given, that only one
      s-wave resonance is not sufficient, the  S$_{11}$(1535)  resonance
      accounts only for the part of the $\eta$ production total cross section,
      for the energy range of the S$_{11}$(1535) dominance.
      A single resonance model for the s-wave only
      is,  therefore,  incapable  of describing all of $\eta$ production
      data, but it may suffice for  the  region  of  the  S$_{11}$(1535)
      dominance, which corresponds to $p_{\pi}$ less than 850 MeV/c.  To
      set the stage, we have tested a simple single resonance model  for
      all three dominant partial waves;  S$_{11}$, P$_{11}$ and D$_{13}$
      without any background terms added.  The model describes the large
      near  threshold  peak,  but  it fails miserably in comparison with
      data above $T_{\pi}$ = 800 MeV.  The recent single resonance model
      by  Bennhold  and  Tanabe  \mbox{(B-T)}  \cite{Ben 91} is, anyhow,
      limited to $T_\pi$ less than 700 MeV.  It describes  the  dominant
      peak  pretty  well  on the gross scale, but significantly fails in
      giving details like the exact peak position, etc.  (see Fig.  1b).
      The  reason for the failure is that the B-T analysis relies on the
      data of Ref.  \cite{Bro 79}, which  suffer  from  a  serious  beam
      momentum  calibration error \cite{Cla 92}.  We have also used data
      of Ref.  \cite{Bro 79}, but with caution, as elaborated in  detail
      later in the text.  The comparison between the two models is shown
      in Fig.  1a and in the expanded scale in  Fig.   1b.   Considering
      that  the  model  of B-T \cite{Ben 91} used erroneous data of Ref.
      \cite{Bro 79} and has an energy dependent form  factor,  which  is
      not  included  in our study, the agreement between the predictions
      of our single resonance model and the B-T model, is  satisfactory.
      Of  course, both single resonance models miserably fail to fit the
      data above 800 MeV.

      To  obtain  a  better  description  of the input data we propose a
      multiresonance coupled channel  model  with  a  smooth  background
      added   in   a  unitary  way  \cite{Cut  79}  similar  as  in  the
      Karlsruhe-Helsinki partial wave analysis  (K-H  PWA)  of  $\pi  N$
      elastic  scattering \cite{Hoh 83}.  A data base consisting of $\pi
      N$ K-H PWA of Ref.  \cite{Hoh 83}, with the addition of total  and
      differential $\pi N \rightarrow \eta N$ cross sections
 \cite{Pdg 92,Bro 79,Dei 69,Bul 69,Ric 70,Nel 72,Bin 73,Fel 75,Cha 75}
      has been chosen in order to perform the fitting procedure  of  the
      $\pi  N$ elastic partial waves up to $T_{\pi}$ = 2727 MeV with the
      additional weighting factors based on the  analyses  of  the  data
      reliability  by Clajus and Nefkens \cite{Cla 92}.  Results for the
      obtained resonance parameters slightly deviate from  the  $\pi  N$
      elastic case, but they give acceptable \mbox{$\pi N \rightarrow \eta N$}
      partial wave T-matrices  and  at  the  same  time  predict  the  T
      matrices  for  the  $\eta  N$ elastic process.  Manley and Salesky
      (M-S) \cite{Man 92} have used a  multichannel  and  multiresonance
      K-matrix  approach  to  the  coupled  channel  inelastic  $\pi  N$
      scattering with the main inelastic channel  beeing  the  continuum
      production.   The  $\eta$  production process is only added in two
      partial waves, the S$_{11}$ and F$_{17}$.  The  only  purpose  for
      including  the  $\eta$ production channel is to maintain unitarity
      when the observed loss of flux can not be attributed to any  other
      process.   The  so  called  "$\eta$  production  channel"  have to
      account for ALL missing inelasticity within the partial  wave  and
      cannot  be  related  just  to  $\eta$  production.  Therefore, the
      prediction for the two $\eta$ production partial waves  T-matrices
      S$_{11}$  and  F$_{17}$,  which  come from the M-S model \cite{Man
      92,Man 93} should be used with care.  As the complete study of the
      $\eta$  production process requires all partial waves we have made
      a multiresonance three coupled  channel  fit  with  the  $\eta  N$
      channel  explicitly  included  and  used  as  a  constraint.   The
      multiresonance M-S model \cite{Man 92,Man 93} is complementary  to
      ours  in  the  sense  that our results for the resonance branching
      ratios for the $\pi^2 N$ channel can be compared to the  continuum
      production  branching  ratios  in  M-S model, and they should (but
      only roughly) correspond.
\section{Formulation of the model}
\subsection{The  three body coupled channel formalism}
      The $\pi N \rightarrow \eta N$ process is given by the invariant
      amplitude \\
      $A(W,\cos\theta^*) + \not\! q_\eta B(W,\cos\theta^*)$ \\
      with the standard on-shell partial waves  decomposition  of  $A$
      and $B$:
\bee
   A(W,\cos\theta^*) & =& \frac{4\pi}{\sqrt{q_\pi^{*3} q_\eta^{*3}}}
    \left\{   \hspace{4.0mm}   \sum_{l=0}^{\infty}   T_{l^+}    \left[
    \hspace{4.0mm}
    \sqrt{(E^*_i+m)(E^*_f+m)}(W-m)P_{l}^\prime(\cos\theta^*) \right.
      \right.  \hspace{5.0mm} \nonumber \\
        & & \left.  \left.
      + \sqrt{(E^*_i-m)(E^*_f-m)}(W+m)P_{l+1}^\prime(\cos\theta^*)
       \right] \right. \hspace{3.0mm}   \nonumber \\
      & &   \left.   - \sum_{l=1}^{\infty} T_{l^-} \left[
      \hspace{1.0mm}
      \sqrt{(E^*_i+m)(E^*_f+m)}(W-m)P_{l+1}^\prime(\cos\theta^*)
      \right.  \right.  \hspace{5.0mm} \nonumber \\
      & & \left. \mbox{\rule[-0.5cm]{0cm}{1.0cm}}  \left.
        + \sqrt{(E^*_i-m)(E^*_f-m)}(W+m)P_{l}^\prime(\cos\theta^*)
   \right] \right\}                     \nonumber \\
        & &                                           \\
  B(W,\cos\theta^*) & = & \frac{4\pi}{\sqrt{q_\pi^{*3} q_\eta^{*3}}}
     \left\{ - \sum_{l=0}^{\infty} T_{l^+} \left[
    \hspace{4.0mm} \sqrt{(E^*_i+m)(E^*_f+m)}P_{l}^\prime(\cos\theta^*)
      \right.  \right.  \hspace{5.0mm} \nonumber \\
      & & \left.  \left.
            - \sqrt{(E^*_i-m)(E^*_f-m)}P_{l+1}^\prime(\cos\theta^*)
       \right] \right. \hspace{3.0mm}   \nonumber \\
      & &  \left.   + \sum_{l=1}^{\infty} T_{l^-} \left[
      \hspace{1.0mm}
      \sqrt{(E^*_i+m)(E^*_f+m)}P_{l+1}^\prime(\cos\theta^*)
       \right.  \right. \hspace{5.0mm}  \nonumber \\
      & & \left. \mbox{\rule[-0.5cm]{0cm}{1.0cm}}  \left.
        - \sqrt{(E^*_i-m)(E^*_f-m)}P_{l}^\prime(\cos\theta^*)
   \right] \right\} .                    \nonumber
\eee

    $W$ is the total c.m. energy,

    $\theta^*$ is the c.m. scattering angle,

    $q^*_\pi$ and $q^*_\eta$ are the initial pion and final $\eta$
        c.m. momenta,

    $E^*_i$ and $E^*_f$ are the initial and final nucleon
        c.m. energies,

    $P_l^\prime(z)$ are derivatives of Legendre polynomials,

    $T_{l^\pm}$ are the $\pi N \rightarrow \eta N$ T-matrices,

    and m is the nucleon mass. \\

      The  $\pi  N  \rightarrow  \eta  N$ $T_{l^{+},l^{-}}$ matrices are
      matrix elements of the three channel partial wave T$^{JL}$  matrix
      which is given as:  \\

      \[
         T^{JL} =
         \left(
            \begin{array} {ccc}
      T^{JL}_{\pi\pi} & T^{JL}_{\pi\eta} & T^{JL}_{\pi \pi^2} \\
      T^{JL}_{\eta\pi} & T^{JL}_{\eta\eta} & T^{JL}_{\eta \pi^2} \\
      T^{JL}_{\pi^2\pi} & T^{JL}_{\pi^2\eta} & T^{JL}_{\pi^2\pi^2} \\
            \end{array}
         \right)
      \]
      where various channels are denoted by the index $\pi$  for  $\pi
      N$,  $\eta$  for  $\eta  N$  and  $\pi^2$ for all other channels
      ($\pi\Delta$, $\rho N$, $\pi\pi N$, ...).  The third  channel  is
      effectively  described  as  a  two  body  process  $\pi^2N$ with
      $\pi^2$ being a quasiparticle with a different mass  chosen  for
      each  partial  wave.  We have fixed the channel masses, for each
      partial wave independently.
\subsection{Single resonance model}
      The  simplest  possible model to represent the $\pi N$ vertex, and
      which is  directly comparable to earlier analyses \cite{Bha 85,Chi
      91,Ben  91,Hai  86},  has  been  constructed.   Only one resonance
      without any background terms is used to describe each of the three
      important partial waves S$_{11}$, P$_{11}$ and D$_{13}$.

      The elastic T-matrix for each resonance is defined as:
      \be
      T^{JL}_{cc}(W) = \frac{\Gamma^{JL}_c(W)/2}
                         {M^{JL} - W - i\Gamma^{JL}_{\rm tot}(W)/2};
                         \hspace{1.0cm} c = \pi, \eta, \pi^2.
      \ee
      The partial widths are given by:
      \be
          \Gamma^{JL}_c(W) = \Gamma^{JL}_c(M^{JL}) \times
          \left\{
          \begin{array} {cl}
               \left( \frac{q_c}{q^{JL}_{0c}} \right)^{2L+1}
                & $for$ \  q_c < q^{JL}_{0c}          \vspace{3mm} \\
               \left( \frac{2q_c}{q_c + q^{JL}_{0c}} \right)^{2L+1}
                & $for \ $ q_c > q^{JL}_{0c}
          \end{array}
          \right.
      \ee
    and $q_c$ is the c.m. momentum of the channel meson $c$: $\pi$, $\eta$,
      or $\pi^2$:
      \be
         q_c \equiv q_c(W) =
         \frac{\sqrt{\left(W^2-(m+m_c)^2\right)
                     \left(W^2-(m-m_c)^2\right)}} {2W}
      \ee
      for $m_c$ = $m_\pi$, $m_\eta$, $m_{\pi^2}$.
        $q^{JL}_{0c} \equiv q_c(M^{JL})$ is the c.m. momentum of the
        channel meson $c$ at the resonance mass $M^{JL}$.

      The exception is P$_{11}$ because its mass is below $\eta$
      threshold. For that case we use:
      \be
          \Gamma^{P_{11}}_{\eta}(W) =
          \left\{
          \begin{array} {cl}
               0 &\ \ {\rm below}\ \eta N\ {\rm threshold} \\
               100 \left(\frac{q_\eta}{410}\right)^3
                 &\ \ {\rm above}\ \eta N\ {\rm threshold}
          \end{array}
          \right.
      \ee
      This  definition  of  partial  widths  gives the correct threshold
      behavior for the T-matrix.  The total widths are given as:
      \be
         \Gamma^{JL}_{\rm tot}(W) = \Gamma^{JL}_{\pi}(W)
                              + \Gamma^{JL}_{\eta}(W)
                              + \Gamma^{JL}_{\pi^2}(W).
      \ee
      The inelastic T-matrix is given by
      \be
      T^{JL}_{c_1c_2}(W) = \sqrt{T^{JL}_{c_1}(W) T^{JL}_{c_1}(W)}
                         \hspace{1.0cm} c_1, c_2 = \pi, \eta, \pi^2,
      \ee
      and the T-matrix is defined in such a way as to give a  unitary  S
      matrix:
      \be
          S = 1 + 2iT,\ \ \ \ S^{\dagger}S=SS^{\dagger}=1. \nonumber
      \ee
      Reliable  branching  ratios  of  various  resonances to the $\eta$
      channel unfortunately are not available except for  S$_{11}$.   As
      can  be  seen from Table 1, all values used in our calculation are
      within range of the accepted values \cite{Pdg 92}.   However,  the
      agreement  of  the  elastic $\pi N$ partial wave T-matrices values
      with the Karlsruhe - Helsinki phase shift analysis (K-H) \cite{Hoh
      83}  is  not  entirely  satisfactory  for  the  lower energy range
      $T_{\pi} \leq$ 700 MeV, and the single resonance model  completely
      misses  all  higher  energy partial waves in the elastic channels.
      The results are shown in Figs.  2a and 2b.  The dashed-dotted line
      shows  our  single  resonance model and the dotted line is the B-T
      single resonance  model  of  Ref.   \cite{Ben  91}  which  has  an
      additional energy dependent form factor.  The B-T model is limited
      to lower energies.  The dots are the elastic K-H $\pi  N$  partial
      wave  analyses  \cite{Hoh  83}.   As the K-H PWA does not give the
      error analysis for the partial wave T-matrices in  \cite{Hoh  83},
      and  the  errors are essential to define the statistical weight of
      the analyses, we have identified the errors of the  used  data  in
      the standard $\chi^2$ analysis as:
      \begin{eqnarray*}
      \Delta^{JL}_i &=& 0.005 + \left( 0.01 + 0.0015 \ \frac{W_i-W^{\pi
                                \ {\rm thresh}} }
        { \Delta} \right) |T^{JL}_{\rm max}| \\
               \Delta &=& 1 \ {\rm GeV} \\
              W_i & {\rm is} & {\rm the  \ total \ c.m. \ energy } \\
              W^{\pi \ {\rm thresh}}& {\rm is} & {\rm the \ total \ energy
              \ at \ \pi
             \ nucleon \ threshold}  \\
             |T^{JL}_{\rm max}| & {\rm is} &
                    {\rm the \ maximal \ value \ of \ the \ T-matrix \
                    in \ the \ chosen \ energy \ range.}
      \end{eqnarray*}
      The energy range extents up to 2.5 GeV of the total c.m. energy.

      The statistical weight in the $\chi^2$ function is defined in a
      standard way as well:
      \begin{eqnarray*}
          w^{JL}_i=\frac{1}{(\Delta^{JL}_i)^2}.
      \end{eqnarray*}

      The introduced energy dependence  of  the  statistical  weight  is
      inspired  by  the  energy dependence of the error analysis of Ref.
      \cite{Cut 79}.  It steadily  raises  with  energy,  but  does  not
      exceed the value of 0.02 in the units of Ref. \cite{Hoh 83}.

      The total cross sections for the $\eta$  production  are  compared
      with  the  single  resonance  model  (dash-dotted  line)  of  this
      article, and the B-T single resonance model of Ref.  \cite{Ben 91}
      (dotted  line) in Figs.  1a and 1b.  The experimental data for the
      $\eta$ production differential cross sections  are  compared  with
      both  models  in Figs.  3a-3d.  The full and dashed lines in Figs.
      1-3 will be explained later.

      The failure of single resonance model to describe the experimental
      facts is not unexpected.  Even at the lower  energies,  background
      terms  which are not included in the single resonance model are of
      some importance, and their omission can account for  the  observed
      discrepancy  with the input.  The comparison of the parameter free
      single resonance model, of this article (dash-dotted  line),  with
      the  similar  B-T  model  \cite{Ben 91} (dotted line) shows a fair
      likeness in the energy range where the  later  analysis  has  been
      given.   The  differences between the two models are attributed to
      the different  input  data  base  and  to  the  additional  energy
      dependent  form factor used in \cite{Ben 91}.  However, as is seen
      in Figs.  1-3, both models  show  reasonable  agreement  with  the
      input  data  for  the  important  $\pi  N$  elastic  partial waves
      S$_{11}$, P$_{11}$ and D$_{13}$ at lower energies.   Our  analysis
      predicts strong deviations from the fitted $\pi N$ elastic partial
      waves at  higher  energies,  for  the  S$_{11}$  partial  wave  in
      particular,  so it is to be expected that the results of \cite{Ben
      91} suffer from a similar disease for  higher  energies.   Similar
      statements  can  be  and  are valid for the total and differential
      cross sections  of  $\eta$  production  (Figs.   1  and  3).   The
      phenomenon  observed  is  consistent with the fact that the single
      resonance model covers only the energy range of the first peak  in
      $\sigma_t(\pi p \rightarrow \eta n$).

      Let us mention that in the described single resonance formalism it
      is not possible to extrapolate the model to include more than  one
      resonance  per  partial  wave  in a straightforward manner without
      directly violating the S-matrix unitarity.
 \subsection{  A unitary multiresonance model}
      In  order to fit the $\pi N$ elastic amplitudes of Ref.  \cite{Hoh
      83} better then in single resonance model  we  have  introduced  a
      manifestly  unitary  model  that  enables  including more than one
      resonance and background term per partial wave.  It is constructed
      following  the  commonly  accepted  method  developed in \cite{Cut
      79}, and originally used in $\pi N$ elastic partial  wave
      analysis (PWA) \cite{Hoh 83}.  \footnote{The simple recipe for the
      modification of the S-matrix, \vspace*{-0.3cm} which  consists  in
      directly  adding  extra  resonances  and  a  smooth  background is
      manifestly nonunitary.}

      As  the $\eta $ meson is a pseudoscalar, isospin zero particle, it
      mixes by isospin violation with the $\pi^0$.  We have  chosen  the
      following  three coupled channels to set up the model:  the $\pi N
      $, $ \eta N$ and a third, an effective two  body  channel  labeled
      $\pi^2   N$,   which   inclusively  contains  and  represents  all
      remaining, even three body channels ($\pi \Delta$, $\rho N$,  $\pi
      \pi   N$,   etc.).    The   objective   of  the  procedure  is  to
      simultaneously achieve a good representation of the input $\pi  N$
      elastic  T-matrices,  and  the experimental $\eta$ production data
      (total and differential cross sections) by the values  coming  out
      of the model.

      We distinguish three basic steps of  the presented analysis. \\
      {\bf Step 1 - formalism} \\
      The  multichannel  $T$  matrix taken over from  \cite{Cut 79}
       is given as:
      \be
      T^{JL}_{ab} = \sum_{i,j=1}^{N^{JL}} f^{JL}_a(s) \sqrt{\rho_a}
    \gamma^{JL}_{ai}  G^{JL}_{ij}(s)  \gamma^{JL}_{jb}   \sqrt{\rho_b}
     f^{JL}_b(s)
     \label{eq:tab}
      \ee
       $a,b  =   \pi,  \eta,  \pi^2$.  The initial and final
      channels couple through intermediate "particles"  or  resonances
      labeled with $i$ and $j$, and
      \be
       f^{JL}_a(s)  =  \left(
      \frac{q_a}{Q_{1a} + \sqrt{Q_{2a}^2 + q_a^2}} \right)^L
       \ee
      \be
      \rho_a(s) = \frac{q_a}{\sqrt{s}}
      \ee
      where $s = W^2$ and $q_a$ is the meson momentum  for  any  of  the
      three channels given as:
      \be
         q_a \equiv q_a(W) =
         \frac{\sqrt{\left(W^2-(m+m_a)^2\right)
                     \left(W^2-(m-m_a)^2\right)}} {2W}.
      \ee
      The  mass parameter $m_{\pi^2}$ for the $\pi^2 N$ channel is fixed
      prior to minimization to the mass value at which the partial  wave
      inelasticities show     the opening of the first inelastic channel:
 \begin{center}
 \begin{tabular}{ccccccccc}
 \hline
  Partial wave\ \ & S$_{11}$\ \ &P$_{11}$\ \ &P$_{13}$\ \ &D$_{13}$\ \
               & D$_{15}$\ \ &F$_{15}$\ \ &F$_{17}$\ \ &G$_{17}$\ \ \\
      \hline $m_{\pi^2}$ (MeV) & 450 & 380 & 370 & 380 & 400 & 370 & 650
      & 450 \\
\hline
\end{tabular}
\end{center}
      $\gamma^{JL}_{ia}$ are free parameters and will be determined by
      the fitting procedure.  For the $Q_{1a}$ and $Q_{2a}$ parameters
      we choose the values
      \bee
      Q_{1 \pi}  &=& Q_{2 \pi}   = m_{\pi}  \nonumber   \\
      \label{eq:q1q2}
      Q_{1 \eta} &=& Q_{2 \eta}  = m_{\eta}    \\
      Q_{1 \pi^2}&=& Q_{2 \pi^2} = 400\ {\rm MeV}  \nonumber
      \eee
      $G^{JL}_{ij}$  is  a  dressed  propagator  for
      partial wave $JL$ and particles $i$ and $j$;
      \be
    G^{JL}_{ij}(s) =
    G^{0JL}_{ij}(s)    +     \sum_{k,l=1}^{N^{JL}}     G^{0JL}_{ik}(s)
    \Sigma^{JL}_{kl}(s)  G^{JL}_{lj}(s).
    \label{eq:gij}
     \ee
      The bare propagator
      \be
      G^{0JL}_{ij}(s) = \frac{e_i \delta_{ij}}{s_i - s}
      \label{eq:poles}
      \ee
      has  a pole at the real value $s_i$.  The sign $e_i=\pm 1$ must be
      chosen to be positive for poles above the elastic threshold  which
      correspond to resonances.  The nonresonant background is described
      by  a  function  that  consists  of  two   terms   of   the   form
      (\ref{eq:poles}) with pole positions below $\pi N$ threshold.  For
      that  case  signs  of  terms  are  opposite.   The  positive  sign
      correspond  to  the  repulsive  and  negative  to  the  attractive
      potential.  $\Sigma^{JL}_{kl}$ is the  self-energy  term  for  the
      particle propagator:
      \be
      \Sigma^{JL}_{kl}(s) = \sum_{a} \gamma^{JL}_{ka}
      \Phi^{JL}_a(s)  \gamma^{JL}_{la}
      \ee
      The   imaginary   part   of   $\Phi^{JL}_a(s)$  is  the  effective
      phase-space factor for the channel $a$:
      \be
      Im  \Phi^{JL}_a(s)  =  \left[   f^{JL}_a(s)   \right]^2
      \rho_a(s) \equiv F^{JL}_a(s)
      \label{eq:imphi}
      \ee
      The real part of $\Phi^{JL}_a(s)$
      is  calculated  using  a  subtracted   dispersion   relation
    \be
    \Phi^{JL}_a(s)     =     \frac{s-s_0}{\pi}     \int_{s_a}^{\infty}
    \frac{F^{JL}_a(s^\prime)}  {(s^\prime-s)(s^\prime-s_0)}  ds^\prime
    \label{eq:disp}
    \ee
      where $s_a = (m + m_a)^2$.

      The advantage of this approach is that it  manifestly  maintains
      the  $S$  matrix  unitarity  for  any number of resonance and/or
      background terms.  The disadvantage is that  the  connection  of
      the   parameters   $\gamma^{JL}_{ia}$   and   $s_i$   with   the
      conventional     resonance     parameters     $M^{JL}_i$     and
      $\Gamma^{JL}_{ia}$ is not direct \cite{Cut 79}, but has
      to be calculated.  \\
      {\bf Step 2 - data base and fitting procedure} \\
      The  input  parameters  for  the fitting procedure are $s_{i}$ and
      $\gamma_{ia}$ which determine the bare propagator and self  energy
      term     for     the     particle     propagator,     see     Eqs.
      (\ref{eq:q1q2}),(\ref{eq:gij}) and (\ref{eq:poles})  respectively.
      The  parameters  Q$_{1a}$  and  Q$_{2a}$  which  occur in the form
      factor given in Eq.  (\ref{eq:tab}) have been fixed to the mass of
      the  channel  $a$  meson.  The ones subtracted dispersion relation
      given in Eq.   (\ref{eq:disp})  is  solved  numerically  with  the
      subtraction  constant  $s_0=s_a$  and  $\Phi_a^{JL}(s_0)=0$.   The
      stability of the solution  has  been  tested  by  calculating  and
      reproducing  the  initial imaginary part, see Eq.(\ref{eq:imphi}).
      The numerical integration has been  performed  using  the  adapted
      Gaussian  quadrature  method with no significant dependence on the
      number of points.  We should mention that the dispersion  relation
      has  been  calculated  only once, and tabulated for further use to
      save the CPU because the parameters which form the  integrand  are
      not varied in the minimization procedure.  \\
 {\bf Data base} \\
      As  the input data to the minimization procedure  we have used:
      \begin{enumerate}
      \item the K-H partial wave $\pi N$ analyses \cite{Hoh 83}
      \item total cross sections for the $\pi N \rightarrow \eta N$
      process
 \cite{Bro 79,Dei 69,Bul 69,Ric 70,Nel 72,Bin 73,Fel 75,Cha 75,Deb 75}
      \item differential cross sections for the $\pi N \rightarrow
      \eta N$ process
      \cite{Bro 79,Dei 69,Bul 69,Ric 70,Nel 72,Fel 75,Cha 75,Deb 75}
\end{enumerate}
      We  have fitted the $\pi N$ elastic $T$ matrices for eight $I=1/2$
      partial waves:  $S_{11}$, $P_{11}$, $P_{13}$, $D_{13}$,  $D_{15}$,
      $F_{15}$, $F_{17}$ and $G_{17}$ using the phase shift analysis K-H
      \cite{Hoh 83} at 90 energies from threshold  to  2.5  GeV  of  the
      total c.m.  energy.  We have also tested the use of (CMU-LBL) $\pi
      N$ elastic partial wave analyses \cite{Cut 79} without any notable
      differences.   At  the  inception  of  this work the only PWA that
      covered the high energy was K-H \cite{Hoh 83}.  Since then the VPI
      group \cite{Arn 91} has extended its analysis approximately to 2.1
      GeV.  We do not anticipate that the use of VPI PWA would introduce
      any  substantial  changes  into  the conclusions coming out of the
      present analysis.  Anyhow, a new analysis based on the VPI PWA  is
      planned  to  be the subject of a future research.  Finally, let us
      comment that we should have used the  total  set  of  experimental
      data  for  $\pi N$ elastic processes instead of limiting the input
      to the model dependent information coming from PWAs.  However,  we
      have  assumed  that  existing  PWAs  represent the data adequately
      enough so we can avoid an enormous CPU time consumption.

      The  data  for  the  second,  $\eta  N$  channel,  are  the $\pi N
      \rightarrow \eta N$ differential cross sections at 81 energies and
      total $\eta N$ production cross sections at 67 energies
\cite{Bro 79,Dei 69,Bul 69,Ric 70,Nel 72,Bin 73,Fel 75,Cha 75,Deb 75}.
      The  statistical  weights  of  for  the $\eta N$ data, used in the
      minimization procedure, have been based on  the  analysis  of  the
      world  data  give  in  Ref.   \cite{Cla  92}.   In  some cases the
      published statistical weight had  to  be  modified.   Problems  of
      consistency  among  different  measurements  have been extensively
      discussed in \cite{Cla 92}, and  the  discrepancy  for  the  lower
      energies  of  $\eta$ production differential cross section of data
      of Ref.  \cite{Bro 79} has been claimed.  It has  been  argued  in
      Ref.    \cite{Cla   92}  the  data  of  Ref.   \cite{Bro  79}  are
      systematically too low, and that it is due to an error in the beam
      momentum  calibration  which  makes  the  data  at lower energies
      essentially unusable even if  one  tries  to  correct  them  by  a
      momentum  shift.   The direct reason for that is a strong momentum
      dependence  of  the  lab  $\Leftrightarrow$  c.m.   transformation
      Jacobian.  However, at higher energies, it is safe just to perform
      a 4 \% momentum shift downward.  Therefore, the systematic  error,
      in  addition  to the published, statistical one, has been added to
      the questionable data sets reducing their statistical impact.  The
      statistical  weight  of  all  differential  cross sections of Ref.
      \cite{Bro 79} has been reduced by the factor of 2.   In  addition,
      the statistical weight at lowest energies of $W$ = 1511 MeV,  1542
      MeV  and  1571  MeV  has  been  even  more  reduced  so that their
      importance in the analysis is practically eliminated.

      A  systematic  error  of  0.01  mb/sr  has been added to {\em all}
      $\eta$ production differential cross sections.  We have decided to
      do  so  because  of  the  fact  that  the  quoted  errors  in  all
      measurements have been of statistical origin only, so most of  the
      errors have been unrealistically low.

      The data taken at the energy of 1507 MeV in Ref.  \cite{Ric 70} as
      well tend to be too low when the total cross section is calculated
      \cite{Cla  92}  and compared to the "world trend".  Therefore, the
      additional systematic error of 0.04 mb/sr have been added to these
      data.

      For similar general reasons the systematic error of 0.02 mb/sr has
      been added to all the data of Ref.  \cite{Deb 75}.

      All "two star" cross sections, as given in Ref.  \cite{Cla 92} are
      taken with the increased  statistical  factor  of  10,  while  the
      statistical  factor of the remaining cross section data is kept to
      be 1.

      The  direct consequence of our choice of the data base statistical
      weights is seen in Fig.  3a:  Ref.  \cite{Ben 91}  reproduces  the
      data set of Ref.  \cite{Bro 79} at lower energies (  $W$  =  1511,
      1546  and 1571 MeV) while our result tends to reproduce other sets
      much better.  Additional, precise measurements  for  that  process
      are needed in order to eliminate the present uncertainties.  \\
      {\bf Fitting procedure} \\
      We have fitted the  8 above listed lowest $I=1/2$ partial
         waves in the following manner:
         \begin{description}
      \item[a.]
      The number of resonances  and  the  shape  of  the  background  is
      determined  by  the choice of bare propagator and self energy term
      parameters.   We  have  decided  to  start  from  the  number   of
      resonances  in each partial wave as given in \cite{Pdg 92,Hoh 83};
      the background is represented by two resonant functions  with  the
      constraint  that  the  pole  position  is far outside the physical
      region.  Only one resonance in the P$_{13}$ and  F$_{15}$  partial
      waves has been used in our analysis, higher (uncertain) resonances
      have been dropped.  The existence of the second P$_{13}$ resonance
      has  been  recently  suggested, and it was used in the M-S $\pi N$
      analyses \cite{Man 92,Man 93}.
      \item[b.] In general, we need up to two  background  terms.   They
      have  been  numerically  represented as tails of resonances having
      their   pole   positions   far   outside   the   analyzed   energy
      range\footnote{More  than one background term was \vspace*{-0.3cm}
      needed in order to  obtain  some  nontrivial,  but  smooth  energy
      dependent behavior of the background terms.}.
      \item[c.] Error analysis for resonance parameters has been done
      on the basis of MINUIT, imposing the confidence level of 70 \%
      \cite{Min 92}.
         \end{description}
      We  have  used  a  standard  MINUIT program using as much as 132
      parameters in the final run.

      The minimization has been complicated, and a  lot  of  technical
      tricks  had to be used to avoid occurrence of local minima which
      are  hard to handle for a  minimization  with  such a number   of
      parameters.

      The  result  of  the  fitting procedure gives us  a full three
      channel  T-matrix,  with  submatrices  describing   the   $\pi   N
      \rightarrow \eta N$ and $\eta N \rightarrow \eta N $ processes. \\
      {\bf Step 3 - resonance parameters} \\
      The  pole  positions,  resonance  masses  and  widths  have  to be
      obtained  numerically  from  the  full  partial   wave   T-matrix,
      following directly the technique developed in \cite{Cut 79}.
      For the convenience of the reader we shall briefly reproduce the
      essential steps, angular momentum indexes are suppressed.

      The poles of T-matrix  given in Eq.(\ref{eq:tab}) are found solving
      the following equation:
      \be
              \det G^{-1}=0.
      \ee
      The eigenvector of the matrix $H\equiv G^{-1}$ has been  found  at
      the pole position $s_{\rm pole}$:
      \be
         \sum_{j} H_{ij}(s_{\rm pole})\chi_{j}= 0\,.
      \ee
      We have defined quantities:
      \be
        \eta_{c}=\sum_{i} \gamma_{ic} \chi_{i}
      \ee
      which define the coupling of resonance $i$ to the channel $c$.
      We consider the width to be an energy-dependent quantity involving
      the phase-space factors:
      \bee
         \Gamma &\sim& \sum_{c}y_{c} F_{c}(s) \nonumber \\
          y_{c}&=&|\eta_{c}|^2.
      \eee
      Near the resonance we parametrize the T-matrix as:
      \be
        T_{ab}= (B_{ab}-\delta_{ab})/2i +
             \sum_{cd}B_{ac}^{1/2}F_{c}^{1/2}\eta_cD^{-1}
              \eta_{d}F_{d}^{1/2}B_{db}^{1/2}
       \ee
       where $B_{ab}$ is a background S-matrix and  the generalized
       Breit-Wigner denominator is:
      \be
         D(s) = r - s - c \sum_c y_c \Phi_c(s).
         \label{eq:BWD}
      \ee
      The  real  constants  $r$  and  $c$  are  chosen so that $D(s_{\rm
      pole})=0$.  The resonance mass, width and the branching ratios are
      defined as:
      \bee
         Re\ D(M^2) & = & 0     \nonumber \\
         \Gamma   &=& \frac{Im\ D(M^2)} {M Re\ D^\prime (M^2)} \\
         \Gamma_c &=& \frac{y_c F_c(M^2) \Gamma}
                           {\sum_a y_a F_a(M^2)} \equiv x_c \Gamma
                           \nonumber
      \eee
      where $D^\prime$ is the derivative of the generalized Breit-Wigner
      denominator  (\ref{eq:BWD}).   The  obtained  resonance  parameter
      values are given in Tables 2 and 3.
      \section{Results of the unitary multiresonance  model  with  three
                P$_{11}$ resonances}
      We have obtained the partial wave T-matrices for $\pi  N$  elastic
      scattering (see Figs.~2a and 2b), $\pi N \rightarrow \eta N$ (see
      Figs.  4a and 4b) and $\eta N$ elastic scattering (see  Figs.~5a
      and 5b), on the basis of our fit to the $\pi N$ elastic and $\pi N
      \rightarrow \eta N$ data.  The agreement with the  input  $\pi  N$
      elastic  K-H PWA T-matrices is given in Fig.~2, while the agreement
      with the input $\pi  N  \rightarrow  \eta  N$  differential  cross
      sections  is  given  in  Fig.~3.   The full, dash-dotted and dotted
      lines systematically denote the results of our multiresonance, and
      the   two  single  resonance  models,  this  publication  and  B-T
      respectively.  The T-matrix for $\eta N$ elastic scattering  is  a
      prediction.   All  observables,  obtainable  on  the  basis of our
      results for the partial wave T-matrix of the  $\pi  N  \rightarrow
      \eta  N$ and $\eta N$ elastic reactions, with the exception of the
      $\pi N \rightarrow \eta N$ differential cross sections, which  are
      input  data,  are  henceforth a prediction.  Our T matrices can be
      used as input to $\eta$ production calculations in reactions  such
      as $p\,p \rightarrow p\,p\,\eta$.

      As can be  seen,  single  resonance  models  give  a  reasonable
      agreement only in the region of the S$_{11}$(1535).  They can be
      used for higher order calculations only, for  a  limited  energy
      range  where one resonance per partial wave dominates.  However,
      the analyses presented in this article, which is based  on  more
      than  one  resonance per partial wave, can be used over the full
      energy range.

      The   results   of  the  multiresonance  model  for  the  $\pi  N$
      interaction by M-S \cite{Man 92} can  be  used  as  a  consistency
      check.   The  parameters  of  both  models, for the same number of
      resonances per partial wave ( those given  by  the  PDG  \cite{Pdg
      92})   are  listed  in  Table~2.   We  may  call  these  analyses
      complementary because the inelastic part of the  $\pi  N$  partial
      waves  is  constrained  by  two  complementary  processes:  $\eta$
      production in our case, and continuum production in  the  case  of
      Ref.~\cite{Man 92}.  Therefore, the $\pi^2$ part of our analysis
      should roughly  correspond  to  the  $\pi  \pi$  part  of  another
      analyses,  and  the  parameters  of  the $\eta$ production partial
      waves explicitly included in other analyses should  correspond  to
      our  findings.   Of course, masses and widths of resonances should
      correspond to the values given by the  K-H  and  CMU-LBL  $\pi  N$
      elastic  analyses,  which  are generally accepted by PDG \cite{Pdg
      92}.

      As can be seen from our  Figs.~1-3,  the  multiresonance  model,
      based  on  the  standard  number of resonances describes the input
      $\pi N \rightarrow \pi N$ and $\pi  N  \rightarrow  \eta  N$  data
      fairly  well.  Of course, the $\eta$ production cross sections are
      as well  correctly  described  in  the  full  energy  range.   The
      structure  of  the  $\pi N$ elastic partial waves \cite{Hoh 83} is
      not  entirely  reproduced.   The  tendency  of  smoothing  elastic
      partial waves, as it has been already indicated previously, exists
      when the inelastic channels are explicitly  included.   Therefore,
      we  are  tempted to conclude that inclusion of inelastic processes
      imposes some restrictions on the elastic channel  forcing  partial
      waves to have less structure then in \cite{Hoh 83}.

      Comparing the available analyses we conclude:
\begin{enumerate}
      \item  The masses and $\pi N$ elastic branching ratios generally
      agree for all three PWA, with the  exception  of  the  P$_{11}$.
      This problem is discussed later.
      \item   All   partial  waves  {\em  except}  the  P$_{11}$  show
      reasonable agreement in both multichannel models.
      \item The two higher P$_{11}$ partial wave resonances of  M-S  and
      our  analyses  do  not  agree.   The  branching  ratio  for $\eta$
      production in our model is about 90 \%.  The M-S  model  \cite{Man
      92}  predicts  almost  a  60  \%  branching ratio to the $\pi \pi$
      channel, leaving no freedom for  any  flux  going  to  the  $\eta$
      production  channel.   The  disagreement  is obvious, and we shall
      offer a natural explanation.
\end{enumerate}
\section{Results of  the unitary multiresonance model with
      four P$_{11}$ resonances}
      Inspection  of  the  resonance parameters of Ref.  \cite{Man 92}
      reveals the following (see Table 2.):
\begin{itemize}
      \item the total width of the P$_{11}(_{1440/135}^{51})$ is
      different  from the ones of K-H \cite{Hoh 83}, CMU-LBL
      \cite{Cut 79} and Arndt et al.  \cite{Arn 91}
      \item  the total width of the P$_{11}(_{1710/120}^{12})$ is
      different  from the ones of K-H \cite{Hoh 83}, CMU-LBL
      \cite{Cut 79} and Arndt et al.  \cite{Arn 91}
      \item the mass the P$_{11}(_{2100/200}^{9})$ is much lower  than
      the ones of K-H \cite{Hoh 83} and CMU-LBL \cite{Cut 79}
      \item  the mass of the D$_{13}(_{2080/265}^{6})$ is shifted
      from 2080 to 1804 MeV
\end{itemize}
      Therefore, we suspect that a part of the physics in the vicinity
      of the 1800 MeV mass region is not entirely taken into account.

      {\em We assume that there is  another degree of freedom in the
      P$_{11}$ partial wave, in the form of another resonance.}

      This possibility of having four instead of three resonances in the
      P$_{11}$ channel leads to a fit which is shown in the figures with
      dashed lines.  The resonance parameters are given in Table~3.  The
      T-matrix pole positions for K-H  \cite{Hoh  83},  CMU-LBL\cite{Cut
      79},  M-S \cite{Man 92} and our analyses (3 and 4 poles) are given
      in Fig.~6\footnote{CMU-LBL  and  our  analyses  give  the  exact
      T-matrix  pole positions, \vspace*{-0.3cm} while the poles for the
      remaining analyses have been approximated by $M -  i\,\Gamma/2$.}.
      As  can  be  seen  in  Fig.~6,  the  pole positions of the three
      resonances are well established in the classical $\pi  N$  elastic
      PWAs  and  are fairly close.  The P$_{11}$ T-matrix pole positions
      for the M-S analysis \cite{Man 92} (given by filled  circles)  are
      quite  different  from  the  recommended  values.   The three pole
      version  of  the  P$_{11}$  of  our  calculation  is  also   quite
      compatible with the K-H and CMU-LBL pole positions.  The four pole
      version of the P$_{11}$ shows agreement for the lowest  pole,  the
      next two poles are near in masses, but distinctly separated in the
      complex energy plane, while the fourth pole is somewhat lower than
      K-H.   In our analysis the fourth resonance is strongly inelastic,
      going mainly to the $\eta$ production channel, therefore it is not
      to  be  expected  that  either  of  the  K-H or M-S analyses could
      determine it with great precision.
 \section{Conclusions}
 \begin{enumerate}
      \item The addition of another resonance in the P$_{11}$  partial
      wave  definitely  improves our fit to elastic and inelastic data
      in  all  channels.   Various  quark models also predict four and
      even five P$_{11}$ resonances in the energy region  of  1440  to
      2200 MeV \cite{Kon 80,Cap 94}.
      \item The changes in $\pi N$ elastic partial waves are  negligible
      because  all  the resonances (with the exception of the first one)
      are strongly inelastic.
      \item  The  changes  in  $\eta$  production  and  $\eta N$ elastic
      channel T-matrices are clearly visible in all partial  waves.   As
      it is to be expected, the P$_{11}$ is significantly changed, while
      other partial waves do show some variation.  However, let us  draw
      the  readers  attention  to  the  fact that S$_{11}$, P$_{11}$ and
      D$_{13}$  partial  waves  are  the  dominant   ones,   while   the
      contribution  of other channels is at least the order of magnitude
      lower.  So, even if the relative change in Figs.   4  and  5.   is
      large for other partial waves, the change at the absolute scale is
      comparably small.
      \item The inconsistency problem between  the  two  inelastic  PWAs
      goes  away.   The  two P$_{11}$ resonances in the vicinity of 1750
      MeV are responsible for the continuum production, this  is  to  be
      compared with 1717 and 1885 resonances of Ref.  \cite{Man 92}, the
      third resonance at 2215 almost completely couples to the $\eta  N$
      channel,  with  a  very  small  branching  ratio  to the continuum
      production.
      \item  The  M-S  PWA  \cite{Man  92}  could  not  easily  see  the
      additional  fourth  P$_{11}$ resonance as it mostly couples to the
      $\eta$  production  channel,  and  it  is  not,  in  their   case,
      explicitly included.
      \item The elastic $\pi N$ analyses  \cite{Hoh  83,Cut  79}  have
      problems  of  determining  the  number  and  parameters  of  the
      resonances going to inelastic channels.   Some  changes  in  the
      resonances  with  a small elastic branching ratio should be easy
      to obtain in the single channel formalism.
      \item If the Brown data for $T_\pi > 900$  MeV  are  shifted  to
      lower energies by 30 MeV and data below 900 MeV are omitted then
      the resonance parameters for our solutions containing  3  and  4
      resonance in the P$_{11}$ are shown in Table 4.  The results are
      virtually identical (within errors) with the solutions using all
      originally  published  Brown  data \cite{Bro 79}.  That indicates
      that our PWA is showing  robustness  with  respect  to  possible
      errors in input.
\end{enumerate}
      Of course, adding extra P$_{11}$ resonance  is  only  one  way  of
      making  our  analysis compatible to M-S PWA, there might be other,
      equally good explanations of the apparent disagreements  when  the
       inelastic  branching ratios are compared.  One technical
      detail  has  to  be  kept  in  mind:   the  background  parts  are
      represented  by  two  resonances  with  poles kept far outside the
      range of interest.  Therefore, the background is also fitted.  The
      proper  way  should  be  to calculate the background in some model
      (for instance a cloudy bag model ), to fix it, and then  fit  just
      the  resonant  part \cite{Lee 94}.  The interference of background
      and resonant parts might shift the resonance parameters notably.

      The   T  matrices  obtained  in  this  study  are  an  essential
      ingredient   for   calculating   $\eta$   production   reactions
      D($\pi$,NN)$\eta$  and  D($\pi$,NN)$\pi$.   The  off  mass shell
      extrapolation procedure for the $\eta$ production amplitudes can
      not  be  determined  in  this  formalism.   We  do hope to learn
      something about these effects from the higher  order  processes,
      assuming that the off-mass shell behavior of the $\eta$ particle
      is similar as of the pions \cite{Bat 94}.
      \footnote{This work has been partly \vspace*{-0.3cm} supported by
      the EC contract CI1*-CT-91-0894 and DOE  contract DE - FG03 -
      88ER40420/A606.}

\section*{Table captions}
\begin{description}
\item[Table 1.] {\em Resonance parameters of the single resonance
                model.}
\item[Table 2.] {\em Resonance parameters of the multiresonance
      model with 3 $P_{11}$ resonances.}        \\
      The results of elastic $\pi N$ analyses
      \cite{Pdg  92,Hoh 83,Cut 79} are given in the first column.  The
      results of the partial wave analysis of this article, as well as
      results of the PWA of Ref.  \cite{Man 92} are given in columns
      2-10.
\item[Table 3.] {\em Resonance parameters of the multiresonance
             model with 4 $P_{11}$ resonances.}
\item[Table 4.] { \em Resonance parameters of the multiresonance
      models with 3 $P_{11}$ and 4 $P_{11}$ resonances} using the data
      of Ref.  \cite{Bro 79}, for $T_\pi > 900$ MeV  shifted  downward
      by  30  MeV.   Brown data below 900 MeV are omitted from fitting
      procedure.
\end{description}
\section*{Figure captions}
 \begin{description}
      \item[Fig.  1.] The total $\eta$ production cross sections.  The
      experimental  data are taken from the literature, as it has been
      indicated  in  the  figure  itself.   The  dotted  line  is  the
      approximation  that  only  one  resonance has a strong branching
      ratio to the $\eta$ production channel, the  results  are  taken
      from  \cite{Ben  91}.   The dash-dotted line is the result of this
      calculation, and is based on similar assumptions.  The full line
      is  the result of the three coupled channel multiresonance model
      presented in this publication  with  the  number  of  resonances
      given  by  the  PDG  \cite{Pdg 92}, namely three in the P$_{11}$
      partial wave.   The        dashed  line  is  the  four  P$_{11}$
      resonance  model.  The full lines and the      dashed lines show
      notable difference for the energies above $T_\pi \geq 800$  MeV.
      The  four resonance model for P$_{11}$ is closer to experimental
      data.

      Fig. 1a is the entire energy range available.

      Fig. 1b is the same data, with an expanded  scale  for
      $T_\pi$ up to 800 MeV.
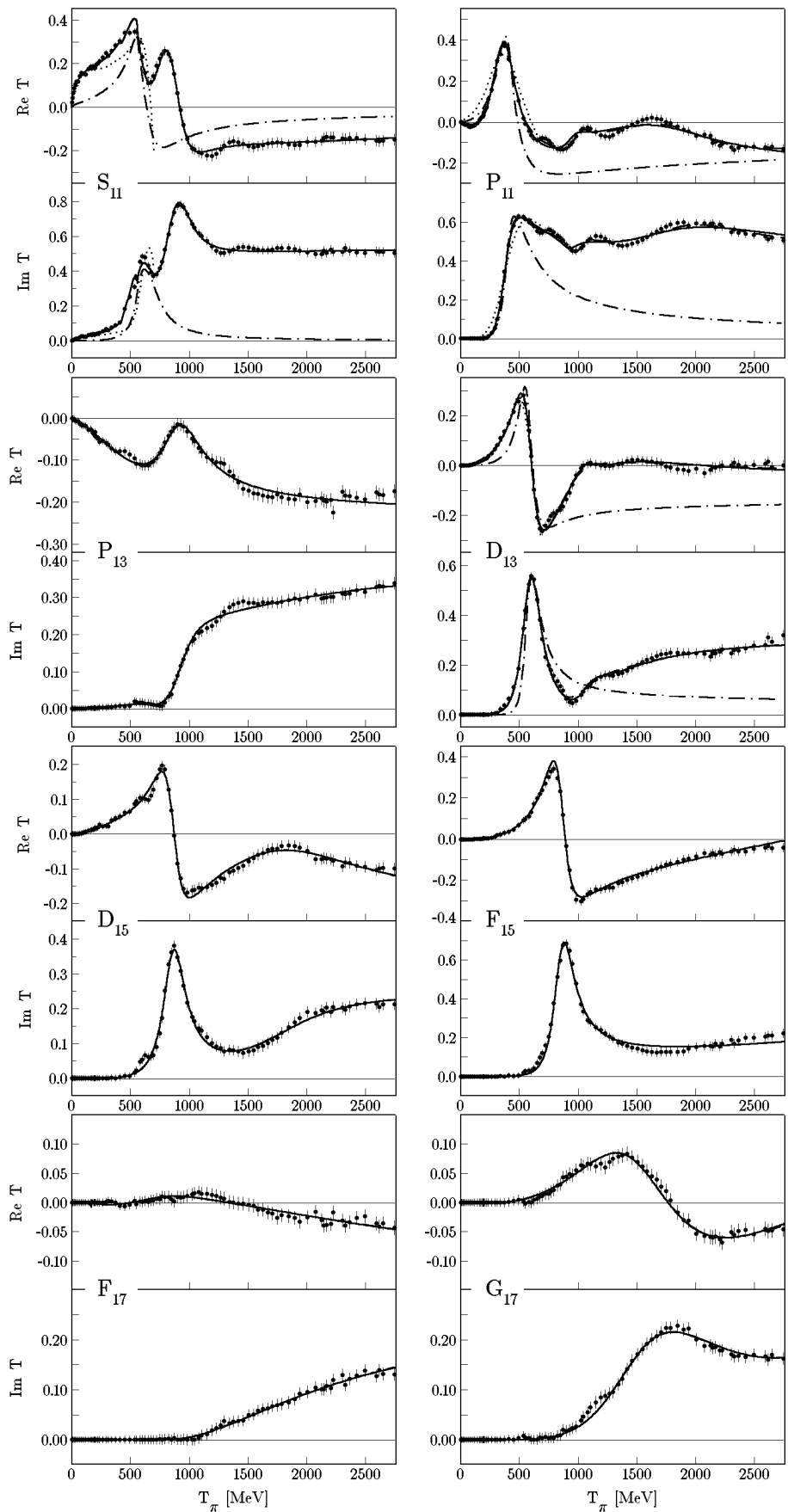
      \item[Fig.   2.]  The  $\pi  N$ elastic partial waves.  The filled
      circles are the result of the single channel $\pi N$  elastic  PWA
      given  in  Ref.   \cite{Hoh  83}.   The used PWA does not give the
      error analyses for the partial wave T-matrices in \cite{Hoh 83}, so
      the  error  bars  given  in the figure are defined in the text and
      reflect the statistical  weight  of  the  data  set  used  in  the
      minimization procedure.  The dotted line is the approximation that
      only one resonance has a strong  branching  ratio  to  the  $\eta$
      production  channel,  the  results  are  from  \cite{Ben 91}.  The
      dash-dotted line is the result of this calculation, and is based on
      similar  assumptions.   The  full  line is the result of the three
      coupled channel  multiresonance  model  presented  here  with  the
      number  of resonances given by the PDG \cite{Pdg 92}, namely three
      in P$_{11}$ partial wave.   The        dashed  line  is  the  four
      P$_{11}$  resonance model.  The full and the      dashed lines are
      practically indistinguishable for all elastic partial waves.
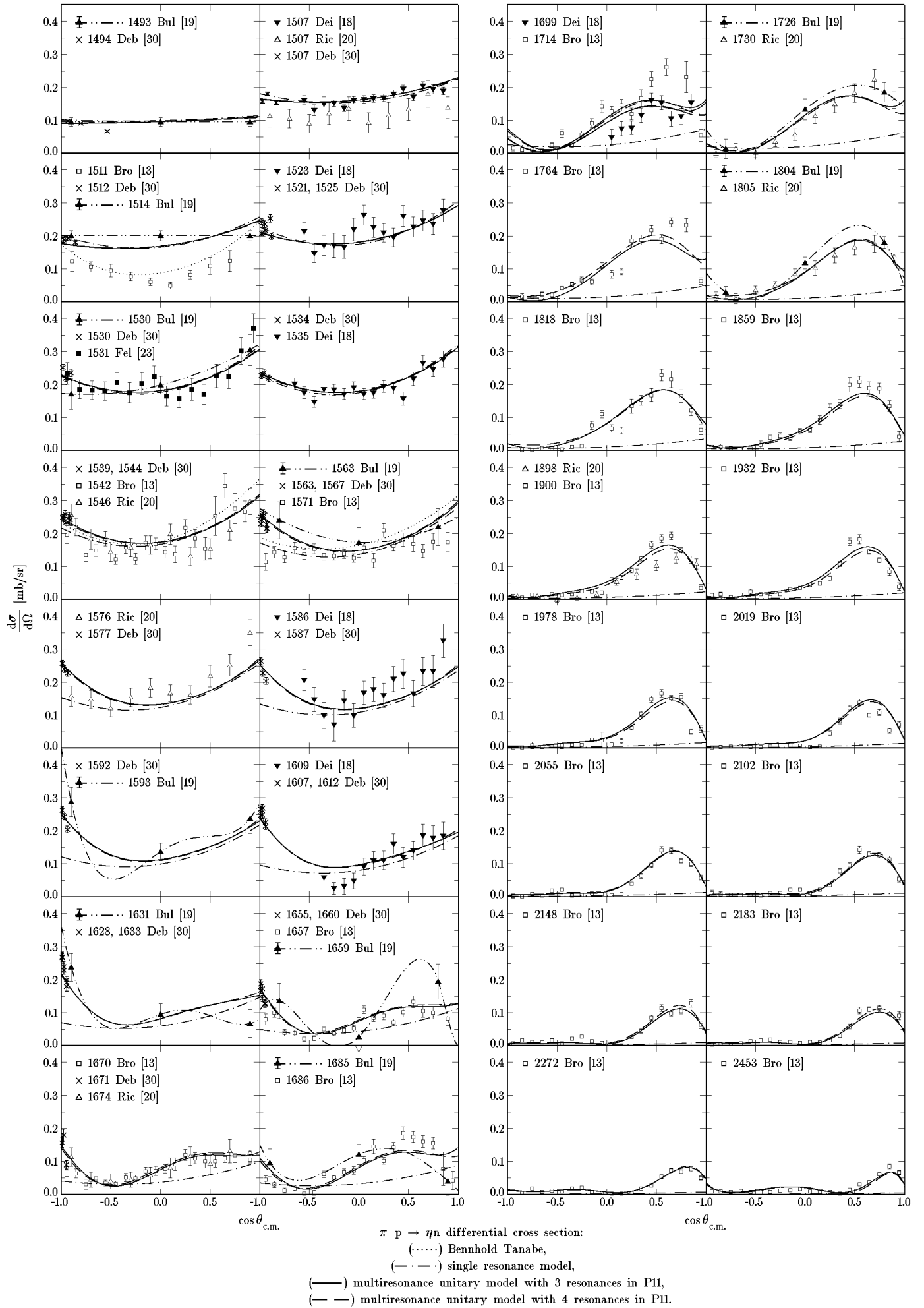
      \item[Fig.    3.]   The  comparison  of  the  $\eta$  production
      differential cross sections with the afore mentioned PWAs.   The
      experimental data  taken from the literature are defined  in the
      figure.  Curves are as in Fig.  2.
      \item[Fig.   4.]  The  partial  wave  T  matrix  for  the   $\eta$
      production  channel.  The dash-dotted line is the single resonance
      approximation for S$_{11}$, P$_{11}$ and D$_{13}$ partial waves of
      this  article.   The  full line is the result of the three coupled
      channel multiresonance model presented here, with  the  number  of
      resonances  given  by  the  PDG  \cite{Pdg  92},  namely  three in
      P$_{11}$ partial  wave.   The  long  dashed  line  represents  the
      increase   to  four  resonances  in  the  P$_{11}$  partial  wave.
      Differences between long dashed and full lines are  quite  notable
      in this figure.
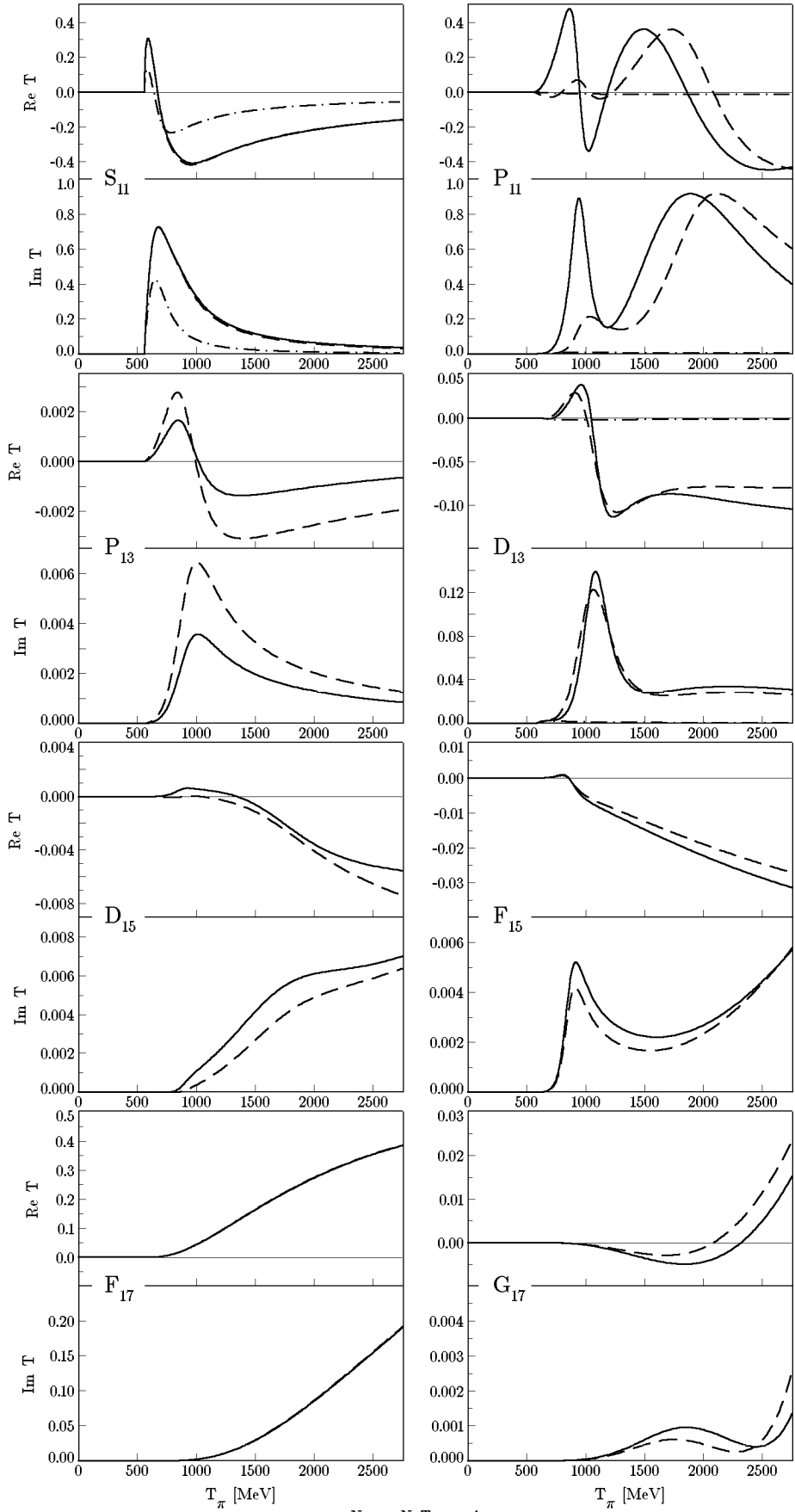
      \item[Fig.  5.] The partial wave  T  matrix  for  the  $\eta  N$
      elastic  channel.   Differences  between       dashed (4 P$_{11}$
      resonances) and full lines (3  P$_{11}$  resonances)  are  quite
      pronounced.
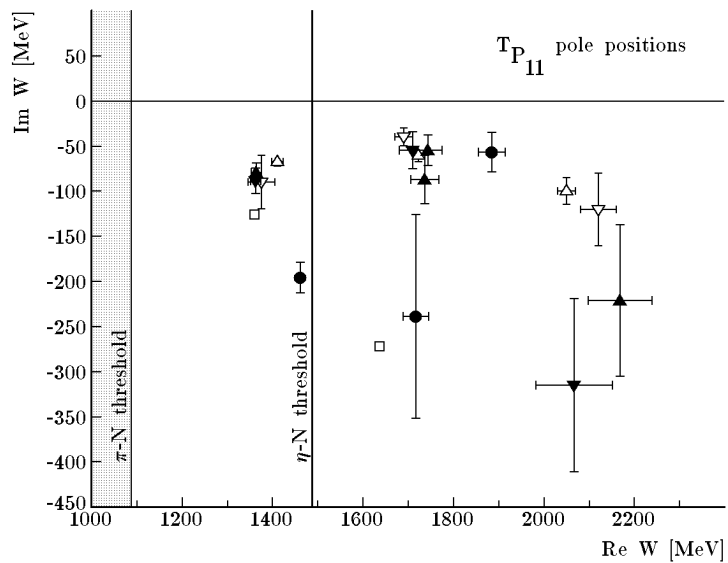
      \item[Fig.   6.]  The  pole  positions for all four partial wave
      analyses, mentioned in this  article.   The  notation  is  given
      explicitly in the figure.
\end{description}
\newpage
\begin{center}
      {\large \bf Table \vspace{1.cm} 1}  \\
\begin{tabular}  {ccc}                            \hline\hline
 Resonance & P.D.G parameters    & Used parameters   \\ \hline\hline
 S11(1535) &    m = 1520 to 1555       &   1535      \\
           & $\Gamma$ = 100 to 250     &    150      \\
           & $x_{\pi}$ = 0.35 to 0.55  &   0.40      \\
           & $x_{\eta}$ = 0.30 to 0.50 &   0.42      \\ \hline
 P11(1440) &     m = 1430 to 1470      &   1440      \\
           & $\Gamma$ = 250 to 450     &    250      \\
           & $x_{\pi}$  = 0.60 to 0.70 &   0.60      \\
           & $x_{\eta}$ = 0            &  $0^{*)}$   \\ \hline
 D13(1520) &     m = 1515 to 1530      &   1520      \\
           & $\Gamma$ = 110 to 135     &    120      \\
           & $x_{\pi}$ = 0.50 to 0.60  &   0.55      \\
           & $x_{\eta} \sim$ 0.001     &   0.0015     \\ \hline\hline
\end{tabular}
\\
\vspace{1.5mm}
$^{*)} x_{\eta} = 0 $ i.e. $ \Gamma_{\eta} = 0 $ is used below,
      $  \Gamma_{\eta}  =  100 (\frac{q_\eta}{410})^3 $ above
      $ \eta N $ threshold (see text).
\end{center}
\end{document}